# Bayesian Structural Identification using Gaussian Process Discrepancy Models

Antonina M. Kosikova[1], Omid Sedehi[2], Costas Papadimitriou[3], Lambros S. Katafygiotis[4,*]

**Abstract**
Bayesian model updating based on Gaussian Process (GP) models has received attention in recent years, which incorporates kernel-based GPs to provide enhanced fidelity response predictions. Although most kernel functions provide high fitting accuracy in the training data set, their out-of-sample predictions can be highly inaccurate. This paper investigates this problem by reformulating the problem on a consistent probabilistic foundation, reviewing common choices of kernel covariance functions, and proposing a new Bayesian model selection for kernel function selection, aiming to create a balance between fitting accuracy, generalizability, and model parsimony. Computational aspects are addressed via Laplace approximation and sampling techniques, providing detailed algorithms and strategies. Numerical and experimental examples are included to demonstrate the accuracy and robustness of the proposed framework. As a result, an exponential-trigonometric covariance function is characterized and justified based on the Bayesian model selection approach and observations of the sample autocorrelation function of the response discrepancies.

## 1. Introduction

Finite Element (FE) models of structures are modern tools for structural analysis, design, control, and monitoring. However, the predicting properties of these numerical models depend squarely on the underlying set of assumptions and parameters considered by simulation experts. This issue brings to light updating FE models based on vibration data, obtained from a limited number of sensors optimally placed at specific positions of structures. In the literature, model updating methods are classified as deterministic, probabilistic, and non-probabilistic techniques [1–4], but the focus of this study is on probabilistic techniques.

Bayesian model inference is a general probabilistic tool, which describes the discrepancy between the measured and model responses through a class of probability distributions known as the likelihood function. Then, it updates the user-defined prior distribution of the structural parameters and characterizes a posterior probability distribution [3,5–7]. This distribution represents the uncertainty associated with the structural parameters, indicating the relative plausibility of structural parameters conditional on the modeling assumptions and the data. When multiple candidates of structural models are available, the framework can be extended to quantify the posterior probability of each model candidate and select the best model class [8–11]. In such cases, it is also possible to average over the response of different models and reduce the sensitivity of response predictions to the choice of structural model classes [12,13].

Bayesian structural model updating methods have significantly progressed over the last three decades. Probabilistic identification of model parameters based on modal statistical information has been an active research direction over the last decade [14–20]. Identification of systematic nonlinearities has recently been addressed through Bayesian filtering techniques, e.g., Kalman and Particle Filters [21–26]. Spatial sparsity of structural damage has been incorporated through parameterized prior distributions [27–29]. Test-to-test variability of structural parameters is accounted for through hierarchical Bayesian models, whereby the environmental and operational effects are characterized [30–

[1] Ph.D. Student, Department of Civil and Environmental Engineering, The Hong Kong University of Science and Technology, Hong Kong, Email: akosikova@connect.ust.hk
[2] Postdoctoral Fellow, Department of Civil and Environmental Engineering, The Hong Kong University of Science and Technology, Hong Kong, Email: osedehi@connect.ust.hk
[3] Professor, Department of Mechanical Engineering, University of Thessaly, Volos, Greece, Email: costasp@uth.gr
[4] Professor, Department of Civil and Environmental Engineering, The Hong Kong University of Science and Technology, Hong Kong, Email: katafygiotis.lambros@gmail.com **(Corresponding Author)**

32]. In general, improving the robustness and accuracy of probabilistic models has been an appealing and demanding research direction, which appears to remain at research spotlight for the years to come.

Treatment of data points as statistically independent and Gaussian distributed constructs the most basic mathematical form of the likelihood function. Due to the maximum-entropy principle, the Gaussian distribution is an optimal choice when the probability density function (PDF) is assumed to have fixed second-moment central statistical moments [33,34]. Although this probability model is intended to comply with the principle of model parsimony [8], it turns out to be a too strong assumption in most practical cases as it entirely ignores the correlation and dependency between the measured data points. This issue is particularly the case in handling time-history data sets where the Gaussian White Noise (GWN) assumption appears to be largely violated. Thus, the essence of incorporating the correlation of data points has been noticed even in the first generation of structural identification methods. The exact covariance matrix was derived in [35,36], where the spatial-temporal correlation of data points is characterized by considering the input excitations to be band-limited Gaussian noise. However, due to computational difficulties in performing inference based on exact covariance matrices, it was proposed to substitute them with sample auto-covariance functions curtailed at a sufficiently large length of the signal, essentially longer than a few fundamental period of the linear system of interest [37,38]. Nevertheless, the sample auto-covariance function might cause overfitting due to incorporating an excessively large number of correlation coefficients driven by the data. Recently, auto-Regressive (AR) models were employed for parameterizing the temporal correlation of prediction errors, but these models are not adequately flexible for modeling different mathematical functions [39,40].

Gaussian Process (GP) models offer indispensable tools for considering spatial-temporal correlations through kernel covariance matrices [41]. Kennedy and O'Hagan [42] developed a Bayesian framework for the calibration of computer models wherein the discrepancy between the measured and model outputs are described by GP models. In recent years, various investigators have applied this framework to structural model updating problems, e.g., [43–45]. GP models have also been used extensively for describing the correlation of structural parameters with environmental factors [46–48], estimating input forces [49], developing Kriging surrogate models [13,50], and optimizing sensor configuration [51]. In these works, Squared-Exponential (SE) kernel function has been the default choice to describe the error characteristics, but the SE kernel might not generalize well, especially when error processes appear as periodic patterns. A recent study has employed the exponential-sinusoidal kernel function to account for the periodicity of prediction errors [52]. However, a more natural approach is to choose the best kernel covariance function through Bayesian model selection techniques, as demonstrated in [53]. In a recent study, the exact covariance function of linear systems is used, exhibiting superior prediction performance compared to other kernels [54]. Nonetheless, such a kernel function only holds when the input is Gaussian noise.

Despite these advances, a kernel covariance function customized for model updating problems is missing from the structural dynamics literature. Therefore, in the present paper, structural model updating is reformulated through its augmentation with GP models. Conventional choices of kernel functions are reviewed, and a new prediction-based Bayesian model class selection is proposed to rank different kernel structures. Computational aspects are discussed, providing the detailed flow of steps when Laplace approximation and sampling techniques are employed. Finally, numerical and experimental examples are provided for demonstrating the proposed framework. It is shown that the prediction errors are characterized best via an exponential-trigonometric kernel covariance function, which offers remarkable predictability and generalizability among other choices explored herein.

The paper continues with Section 2, describing the Bayesian formulation proposed for model inference, response predictions, and kernel selection. Section 3 formalizes the computational aspects of this framework and discusses potential difficulties. Sections 4-6 demonstrate the GP framework via numerical and experimental examples. Section 7 summarizes conclusions of this work.

## 2. Proposed Bayesian Methodology
### 2.1. Structural Model Updating

It is desired to update a structural model $\mathbb{M}(\boldsymbol{\theta}) \in M$ and its unknown parameters $\boldsymbol{\theta} \in \mathbb{R}^{N_\theta}$ based on a set of vibration data, where $M$ is a class of structural model. The data $D = \{(\mathbf{x}_i, \mathbf{y}_i) \mid i = 1, 2, ..., n\}$ consists of complete input knowledge subsumed into $\mathbf{X} = [\mathbf{x}_1^T \ \mathbf{x}_2^T \ ... \ \mathbf{x}_n^T]^T$, as well as the incomplete output of the structure $\mathbf{Y} = [\mathbf{y}_1^T \ \mathbf{y}_2^T \ ... \ \mathbf{y}_n^T]^T$, where $\mathbf{x}_i \in \mathbb{R}^{N_x}$ and $\mathbf{y}_i \in \mathbb{R}^{N_o}$ respectively correspond to the input and output of the structure at discrete time instant $t_i = i\Delta t$. When the input $\mathbf{X}$ is fed into the structural model $\mathcal{M}(\boldsymbol{\theta})$, it generates a set of time-history response $\mathbf{f}(\mathbf{X};\boldsymbol{\theta}) = [(f(\mathbf{x}_1;\boldsymbol{\theta}))^T \ ... \ (f(\mathbf{x}_n;\boldsymbol{\theta}))^T]^T$ such that the functional form $f(\mathbf{x}_i;\boldsymbol{\theta}) \in \mathbb{R}^{N_o}$ outputs the model response at time $t_i$ corresponding to those degrees-of-freedom from which the measurements $\mathbf{Y}$ are obtained. Depending on the choice of structural parameters $(\boldsymbol{\theta})$, there exists some discrepancy between the measured and model responses such that the observed response can be written as

$$\mathbf{y}_i = f(\mathbf{x}_i;\boldsymbol{\theta}) + \boldsymbol{\varepsilon}_i(\boldsymbol{\theta}) \tag{1}$$

where $\boldsymbol{\varepsilon}_i(\boldsymbol{\theta}) \in \mathbb{R}^{N_o}$ is a vector of prediction errors corresponding to the time instant $t_i$, described herein through a GP model $\mathbb{P}(\boldsymbol{\varphi}) \in P$ chosen from a class of probability models $(P)$, parameterized by a vector of unknown hyper-parameters $\boldsymbol{\varphi} \in \mathbb{R}^{N_\varphi}$. This probability model is characterized through a kernel covariance function, denoted by $k(\boldsymbol{\zeta}_i, \boldsymbol{\zeta}_i'; \boldsymbol{\varphi})$, where $\boldsymbol{\zeta}_i \in \mathbb{R}^{N_\zeta}$ is a vector of auxiliary variables. The nature of these variables can be chosen based on the information available. These variables are necessary to define the likelihood of observing the response $\mathbf{y}_i$ conditional on $\mathbf{x}_i$ as follows:

$$p(\mathbf{y}_i | \mathbf{x}_i, \boldsymbol{\theta}, \boldsymbol{\varphi}) \sim GP(\mathbf{y}_i | f(\mathbf{x}_i;\boldsymbol{\theta}), k(\boldsymbol{\zeta}_i, \boldsymbol{\zeta}_i'; \boldsymbol{\varphi})) \tag{2}$$

where this GP has the mean vector $f(\mathbf{x}_i;\boldsymbol{\theta})$ and the kernel covariance function $k(\boldsymbol{\zeta}_i, \boldsymbol{\zeta}_i'; \boldsymbol{\varphi})$. This kernel function correlates the samples $i^{th}$ and $j^{th}$ of the prediction errors in the physical space based on their projection into a multi-dimensional Euclidean manifold $\mathcal{S}$, whereon lies the vector of above-mentioned auxiliary variables $\boldsymbol{\zeta}_i \subset \mathcal{S} \in \mathbb{R}^{N_\zeta}$. In this latent manifold, a dot product operation can be defined over $\boldsymbol{\zeta}_i$'s, governing the correlation of data points in the physical space, i.e., $k(\boldsymbol{\zeta}_i, \boldsymbol{\zeta}_i'; \boldsymbol{\varphi}) = \varphi(\boldsymbol{\zeta}_i) \cdot \varphi(\boldsymbol{\zeta}_i')$, where $\varphi(.)$ is a function mapping $\boldsymbol{\zeta}_i$ to $\mathbf{y}_i$ [55]. Given this probability model, the model-updating problem transforms into the identification of the parameters $\boldsymbol{\Theta} = [\boldsymbol{\theta}^T \ \boldsymbol{\varphi}^T]^T$ based on the data. By following a Bayesian strategy, the prior probability model $\mathbb{P}(\boldsymbol{\Theta}) \in U$ is adopted to describe user's initial knowledge $(U)$ about these parameters before updating them based on the data.

The above series of probabilistic reasoning allows establishing a class of probabilistic models $\mathbb{M}_\mathbb{P}(\boldsymbol{\Theta}) \in M_P$, which embodies a composite of the structural model $\mathcal{M}(\boldsymbol{\theta})$, GP model $\mathbb{P}(\boldsymbol{\varphi})$, and prior probability model $\mathbb{P}(\boldsymbol{\Theta})$. Subsequently, the Bayes' rule can be used for updating $\mathbb{M}_\mathbb{P}(\boldsymbol{\Theta})$ as follows:

$$p(\boldsymbol{\theta}, \boldsymbol{\varphi} | \mathbf{Y}, \mathbf{X}, \mathbb{M}_\mathbb{P}) = \frac{p(\mathbf{Y} | \boldsymbol{\theta}, \boldsymbol{\varphi}, \mathbf{X}, \mathbb{M}_\mathbb{P}) p(\boldsymbol{\theta}, \boldsymbol{\varphi} | \mathbb{M}_\mathbb{P})}{p(\mathbf{Y} | \mathbf{X}, \mathbb{M}_\mathbb{P})} \tag{3}$$

where $p(\boldsymbol{\theta}, \boldsymbol{\varphi} | \mathbb{M}_\mathbb{P})$ is the prior PDF; $p(\boldsymbol{\theta}, \boldsymbol{\varphi} | \mathbf{Y}, \mathbf{X}, \mathbb{M}_\mathbb{P})$ is the posterior PDF, which should be calculated; $p(\mathbf{Y} | \mathbf{X}, \mathbb{M}_\mathbb{P})$ is the evidence of a model class, that is, the integral of the numerator over the entire domain of parameters and given by

$$p(\mathbf{Y} | \mathbf{X}, \mathbb{M}_\mathbb{P}) = \int_{\boldsymbol{\varphi}} \int_{\boldsymbol{\theta}} p(\mathbf{Y} | \boldsymbol{\theta}, \boldsymbol{\varphi}, \mathbf{X}, \mathbb{M}_\mathbb{P}) p(\boldsymbol{\theta}, \boldsymbol{\varphi} | \mathbb{M}_\mathbb{P}) d\boldsymbol{\theta} d\boldsymbol{\varphi} \tag{4}$$

where, $p(\mathbf{Y} | \boldsymbol{\theta}, \boldsymbol{\varphi}, \mathbf{X}, \mathbb{M}_\mathbb{P})$ is the likelihood function of all instances of the responses $(\mathbf{Y})$ conditional on the input time history $(\mathbf{X})$, described based on Eq. (2) and given by

$$p(\mathbf{Y} | \mathbf{X}, \boldsymbol{\theta}, \boldsymbol{\varphi}, \mathbb{M}_\mathbb{P}) = N(\mathbf{Y} | \mathbf{f}(\mathbf{X};\boldsymbol{\theta}), \mathbf{K}) \tag{5}$$

where $\mathbf{f}(\mathbf{X};\boldsymbol{\theta})$ is the mean vector, considered equal to the structural model response, and $\mathbf{K} = [k_{ij}]$ is the covariance matrix, described by the kernel covariance function. This Bayesian formulation provides a posterior distribution for the unknown parameters, conditional on the data and the probabilistic model. Unlike the formulation which is mathematically elegant and easy-to-follow, the computation is challenging and cumbersome even in low-dimensional settings as we have to perform an optimization to search for the MPV or run Markov Chain Monte Carlo (MCMC) sampling to draw samples from the posterior distribution. Later, we will elaborate further on the computational aspects of this methodology.

**Remark 1.** The evidence term in Eq. (4) is often difficult-to-calculate, as it would entail an integration over the multi-dimensional space of all parameters. Fortunately, when inferring the unknown parameters, the evidence turns out to be a constant value, which does not govern the MPV or samples of the posterior distribution. Nonetheless, for the model class selection, as will be explained later, its calculation is inevitable.

## 2.2. Response Predictions

After performing Bayesian inference, the next step is to predict unobserved structural response quantities of interest while considering the uncertainty in the structural model parameters, as well as the prediction error model. For this purpose, it is customary to assume that a new set of input data, $\mathbf{X}_{pred} = [\mathbf{x}_{n+1}^T \ \mathbf{x}_{n+2}^T \ ... \ \mathbf{x}_{n+n'}^T]^T$, is available, containing $n'$ data points. This input data set comes from the same input space as the one used for the inference. Due to the likelihood function considered in Eq. (2), the joint distribution of the observed and unobserved responses can be specified as

$$p(\mathbf{Y}, \mathbf{Y}_{pred} | \mathbf{X}, \mathbf{X}_{pred}, \boldsymbol{\theta}, \boldsymbol{\phi}, \mathbb{M}_\mathbb{P}) = N\left(\begin{bmatrix} \mathbf{Y} \\ \mathbf{Y}_{pred} \end{bmatrix} \middle| \begin{bmatrix} \mathbf{f}(\mathbf{X};\boldsymbol{\theta}) \\ \mathbf{f}(\mathbf{X}_{pred};\boldsymbol{\theta}) \end{bmatrix}, \begin{bmatrix} \mathbf{K} & \mathbf{k}_{pred} \\ \mathbf{k}_{pred}^T & \mathbf{K}_{pred} \end{bmatrix}\right) \quad (6)$$

where $\mathbf{Y}_{pred} = [\mathbf{y}_{n+1}^T \ \mathbf{y}_{n+2}^T \ ... \ \mathbf{y}_{n+n'}^T]^T \in \mathbb{R}^{n'N_o}$ is the unobserved response of the structure subjected to $\mathbf{X}_{pred}$ input; $\mathbf{f}(\mathbf{X}_{pred}; \boldsymbol{\theta}) \in \mathbb{R}^{n'N_o}$ is the structural model response in which the input $\mathbf{X}_{pred}$ is replaced; $\mathbf{K}_{pred} \in \mathbb{R}^{n'N_o \times n'N_o}$ is the covariance matrix of the unobserved output calculated based on the kernel function and the axillary parameters $\boldsymbol{\zeta}_{pred}$; $\mathbf{k}_{pred} \in \mathbb{R}^{nN_o \times n'N_o}$ is the cross covariance matrix of $\mathbf{Y}$ and $\mathbf{Y}_{pred}$. From this joint distribution, it is straightforward to derive the distribution of unobserved response conditional on the observed response and the unknown parameters:

$$p(\mathbf{Y}_{pred} | \mathbf{Y}, \mathbf{X}, \mathbf{X}_{pred}, \boldsymbol{\theta}, \boldsymbol{\phi}, \mathbb{M}_\mathbb{P}) = N(\mathbf{Y}_{pred} | \boldsymbol{\mu}_{pred}, \boldsymbol{\Sigma}_{pred}) \quad (7)$$

where $\boldsymbol{\mu}_{pred} \in \mathbb{R}^{n'N_o}$ and $\boldsymbol{\Sigma}_{pred} \in \mathbb{R}^{n'N_o \times n'N_o}$ are the conditional mean and covariance of the response given by

$$\boldsymbol{\mu}_{pred} = \mathbf{f}(\mathbf{X}_{pred}; \boldsymbol{\theta}) + \mathbf{k}_{pred}^T \mathbf{K}_{pred}^{-1} (\mathbf{Y} - \mathbf{f}(\mathbf{X}; \boldsymbol{\theta})) \quad (8)$$

$$\boldsymbol{\Sigma}_{pred} = \mathbf{K}_{pred} - \mathbf{k}_{pred}^T \mathbf{K}^{-1} \mathbf{k}_{pred} \quad (9)$$

By virtue of the principle of total probability, the posterior predictive distribution of the unobserved responses can be calculated from

$$p(\mathbf{Y}_{pred} | \mathbf{Y}, \mathbf{X}, \mathbf{X}_{pred}, \mathbb{M}_\mathbb{P}) = \int_{\boldsymbol{\phi}} \int_{\boldsymbol{\theta}} p(\mathbf{Y}_{pred} | \mathbf{Y}, \mathbf{X}, \mathbf{X}_{pred}, \boldsymbol{\theta}, \boldsymbol{\phi}, \mathbb{M}_\mathbb{P}) p(\boldsymbol{\theta}, \boldsymbol{\phi} | \mathbf{Y}, \mathbf{X}, \mathbb{M}_\mathbb{P}) d\boldsymbol{\theta} d\boldsymbol{\phi} \quad (10)$$

where the expressions inside the integral are obtained earlier in Eqs. (3) and (7), and $p(\mathbf{Y}_{pred} | \mathbf{Y}, \mathbf{X}, \mathbf{X}_{pred}, \mathbb{M}_\mathbb{P})$ is the posterior predictive distribution of the unobserved response. As explained earlier, the integration over the multi-dimensional space of the parameters is computationally challenging, which will be addressed later.

## 2.3. Prediction-based Model Class Selection

The probabilistic framework formulated above can be continued with a Bayesian model class selection by writing another Bayes' rule to evaluate the relative plausibility of the probabilistic model class. Based on [8], one may write:

$$P(\mathbb{M}_\mathbb{P} | \mathbf{X}, \mathbf{Y}) = \frac{p(\mathbf{Y} | \mathbf{X}, \mathbb{M}_\mathbb{P}) P(\mathbb{M}_\mathbb{P})}{\sum_{\mathbb{M}_\mathbb{P} \in M_P} p(\mathbf{Y} | \mathbf{X}, \mathbb{M}_\mathbb{P}) P(\mathbb{M}_\mathbb{P})} \quad (11)$$

where $P(\mathbb{M}_\mathbb{P})$ is the prior probability of the probabilistic model $\mathbb{M}_\mathbb{P}$, $p(\mathbf{Y} | \mathbf{X}, \mathbb{M}_\mathbb{P})$ is the evidence term calculated by Eq. (4), and $P(\mathbb{M}_\mathbb{P} | \mathbf{X}, \mathbf{Y})$ is the posterior probability of $\mathbb{M}_\mathbb{P}$. When the class of probabilistic models ($M_P$) embed different structural models varying in their underlying assumptions while using the same kernel function, this formulation establishes a trade-off between the fitting accuracy and the complexity (in terms of the number of free parameters) of the structural model. However, this formulation does not consider the accuracy of out-of-sample predictions when a kernel covariance function is included. Moreover, in the context of GP models, most conventional kernels provide extremely high accuracy within the training data set, fitting almost exactly to the data except for the measurement noise. Conversely, they do not offer generalizability in terms of predicting dynamical responses.

In this section, we propose a new formulation to consider generalizability of the kernel covariance functions. For this purpose, it is assumed that the posterior distribution of the unknown parameters is obtained from the training data set $D_{train} = \{(\mathbf{X}, \mathbf{Y})\}$ according to Section 2.1. Then, the posterior distribution was used for predicting the held-out data set $D_{pred} = \{(\mathbf{X}_{pred}, \mathbf{Y}_{pred})\}$ according to Section 2.2, where $\mathbf{Y}_{pred}$ is also available but not used for inferring the unknown parameters. In this case, the Bayes' rule provides:

$$P(\mathbb{M}_\mathbb{P} \mid \mathbf{X}, \mathbf{X}_{pred}, \mathbf{Y}, \mathbf{Y}_{pred}) = \frac{p(\mathbf{Y}_{pred} \mid \mathbf{Y}, \mathbf{X}, \mathbf{X}_{pred}, \mathbb{M}_\mathbb{P}) p(\mathbf{Y} \mid \mathbf{X}, \mathbb{M}_\mathbb{P}) P(\mathbb{M}_\mathbb{P})}{\sum_{\mathbb{M}_\mathbb{P} \in M_P} p(\mathbf{Y}_{pred} \mid \mathbf{Y}, \mathbf{X}, \mathbf{X}_{pred}, \mathbb{M}_\mathbb{P}) p(\mathbf{Y} \mid \mathbf{X}, \mathbb{M}_\mathbb{P}) P(\mathbb{M}_\mathbb{P})} \tag{12}$$

where $p(\mathbf{Y} \mid \mathbf{X}, \mathbb{M}_\mathbb{P})$ is the evidence term calculated by Eq. (4), $p(\mathbf{Y}_{pred} \mid \mathbf{Y}, \mathbf{X}, \mathbf{X}_{pred}, \mathbb{M}_\mathbb{P})$ is the posterior predictive distribution given by Eq. (10), and $P(\mathbb{M}_\mathbb{P} \mid \mathbf{X}, \mathbf{X}_{pred}, \mathbf{Y}, \mathbf{Y}_{pred})$ is the posterior probability of $\mathbb{M}_\mathbb{P}$, reflecting the relative plausible of $\mathbb{M}_\mathbb{P}$, conditional on the data and user's judgment. This treatment can also be interpreted as a Bayesian implementation of cross-validation, wherein the measured response history is broken into the training and prediction segments, aiming to consider prediction properties.

When the prior probability $P(\mathbb{M}_\mathbb{P})$ is fixed, the posterior probability of $\mathbb{M}_\mathbb{P}$ will become proportional to the multiplication of $p(\mathbf{Y}_{pred} \mid \mathbf{Y}, \mathbf{X}, \mathbf{X}_{pred}, \mathbb{M}_\mathbb{P})$ and $p(\mathbf{Y} \mid \mathbf{X}, \mathbb{M}_\mathbb{P})$. However, as it might be difficult to compare directly the posterior probability of different model classes, their natural logarithms might be used instead, which leads to

$$\ln P(\mathbb{M}_\mathbb{P} \mid \mathbf{X}, \mathbf{X}_{pred}, \mathbf{Y}, \mathbf{Y}_{pred}) = \ln \underbrace{p(\mathbf{Y}_{pred} \mid \mathbf{Y}, \mathbf{X}, \mathbf{X}_{pred}, \mathbb{M}_\mathbb{P})}_{\text{Posterior Predictive}} + \ln \underbrace{p(\mathbf{Y} \mid \mathbf{X}, \mathbb{M}_\mathbb{P})}_{\text{Evidence}} \tag{13}$$

In this equation, if we ignore the posterior predictive expression, the original formulation presented in Beck and Yuen [8] is recovered. Nevertheless, this new formulation is more general as it enables creating a balance between generalizability (out-of-sample predictions), fitting accuracy in the training data, and model complexity.

**2.4. Kernel Covariance Functions**
The accuracy of GPs depends squarely upon choosing an appropriate kernel covariance function. There can be numerous kernel structures to simulate the desired mathematical function. Thus, it is important to choose the simplest structure that fits into the problem at hand. For selecting the best kernel function, we intend to employ the above model selection methodology. However, in the remainder of this section, we review typical choices of kernel functions and explain under what conditions they might be a legitimate choice.

*2.4.1. Squared Exponential (SE) Function*
The SE, also known as Gaussian kernel or Radial Basis Function (RBF), is perhaps the most popular kernel function [41]. Its properties, such as easy integration with most functions and infinite differentiability, makes it suitable for a variety of problems. This kernel has only two parameters, the variance ($\sigma_f^2$) and the correlation length ($\ell$) that governs the decay rate of correlation. This kernel function can be written as

$$k_{ij} \doteq k(\zeta_i, \zeta_j; \boldsymbol{\phi}) = \sigma_f^2 \exp\left(-\frac{1}{2} \frac{d_{ij}^2}{\ell^2}\right) \tag{14}$$

where $k_{ij}$ is the element $(i,j)$ of the covariance matrix $\mathbf{K}=[k_{ij}]$, $d_{ij}=\left[(\zeta_i-\zeta_j)^T(\zeta_i-\zeta_j)\right]^{1/2}$ is the so-called distance or similarity measure, $\sigma_f^2$ is the variance of individual samples, $\ell^2$ is the length scale used for describing how fast the correlation fades out as the distance measure increases, and $\boldsymbol{\phi}=[\sigma_f^2 \ \ell^{-2}]^T$ is a vector comprising the unknown parameters of the SE kernel function. A schematic view of the kernel function can be seen in Fig. 1(a). Based on this kernel function, training samples with a closer distance to the predicted ones are presumably more informative. Moreover, the SE is also a stationary covariance matrix since it only relies on the distance $d_{ij}=\|\zeta_i-\zeta_j\|$ and not the dynamic values of $\zeta_i$ or $\zeta_j$.

*2.4.2. Periodic Exponential (PE) Function*
For periodic signals, it is reasonable to modify the SE by embedding a sinusoidal function inside the exponential function. By doing so, the PE is expressed as

$$k_{ij} = \sigma_f^2 \exp\left(-\frac{1}{2}\frac{\sin^2(\omega d_{ij})}{\ell^2}\right) \tag{15}$$

In this equation, the coefficient $\omega$ specifies the frequency of the sine function, and $\ell$ is the correlation length. Such a kernel replicates and predicts responses that have the same period as the training data. However, in case of unstable periodicity and non-stationarity of the signal, it fails to predict unobserved responses. This kernel function is displayed in Fig. 1(b), indicating how the correlation decays as $d_{ij}$ grows.

*2.4.3. Multi-Modal Trigonometric Exponential (MMTE) kernel*
A more flexible kernel function can be acquired by multiplying the SE by a cosine function. Since it might be the case that the correlation constitutes more than a single dominant frequency, the kernel function can further be generalized by considering the contribution from $m$ exponential-cosine function, given by

$$k_{ij} = \sum_{k=1}^{m} \sigma_{f,k}^2 \exp\left(-\frac{d_{ij}^2}{\ell_k^2}\right)\cos(\omega_k d_{ij}) \tag{17}$$

where $\sigma_{f,k}^2$ is the amplitude of the correlation, $\ell_k$ is the correlation length of the $k$th cosine function, $\omega_k$ is the frequency of $k$th function. A unimodal example of this kernel is shown in Fig. 1(c). In this kernel function, by limiting the number of contributing/dominant modes up to $m$ modes, one can well characterize an efficient model of mixing sinusoidal functions while retaining the simplicity in terms of the number of parameters. Note that the MMTE function can be regarded as the decomposition of different dynamical modes, resembling an impulse response of a dynamical system. This kernel function is shown to be promising in structural dynamics applications [56], whose theoretical justification is provided in [54].

*2.4.4. Combination with Isotropic Measurement Noise*
An advantageous feature of kernel functions is to create composite kernels that are more flexible. For instance, when the data is polluted by measurement noise, a Gaussian White Noise (GWN) process can be imposed on top of the kernel function of interest to attain additional robustness. In this case, the new kernel can be written as

$$k'_{ij} = k_{ij} + \sigma_n^2 \delta_{ij} \tag{18}$$

where $\delta_{ij}$ is the Kronecker delta function, and $\sigma_n^2$ is the measurement noise variance. Later, we will investigate how these kernel functions perform when applied to structural dynamics problems.

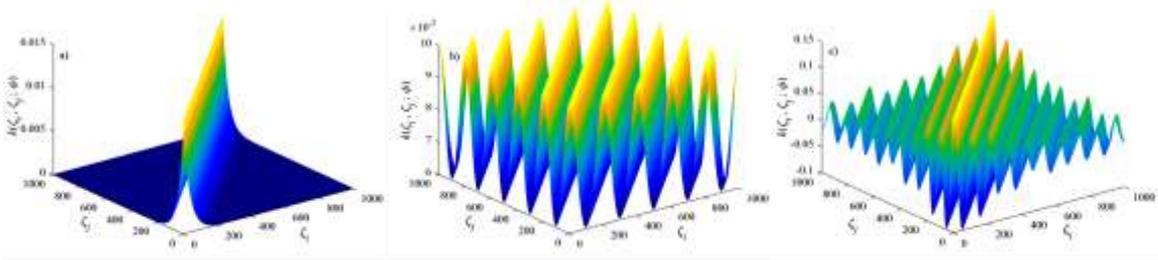

**Fig. 1.** Schematic overview of kernel covariance functions (a) SE (b) PE (c) MMTE

## 3. Computational Methods
### 3.1. Gaussian Approximation of Posterior Distribution

In the presence of a large number of data points, the joint posterior distribution might concentrate around an isolated peak. This condition, referred to as globally identifiable cases [57], allows applying Laplace asymptotic approximation, which substitute the posterior distribution with an approximate Gaussian distribution, centered at the mode of the posterior distribution with a covariance matrix equal to the Hessian matrix inverse evaluated at the mode of the distribution [57]. By applying this approach to Eq. (3), the following approximation can be obtained:

$$p(\boldsymbol{\theta},\boldsymbol{\phi}|\mathbf{Y},\mathbf{X},\mathbb{M}_\mathbb{P}) \approx N\left(\begin{bmatrix}\boldsymbol{\theta}\\\boldsymbol{\phi}\end{bmatrix}\bigg|\begin{bmatrix}\hat{\boldsymbol{\theta}}\\\hat{\boldsymbol{\phi}}\end{bmatrix},\begin{bmatrix}\hat{\boldsymbol{\Sigma}}_{\boldsymbol{\theta}\boldsymbol{\theta}} & \hat{\boldsymbol{\Sigma}}_{\boldsymbol{\theta}\boldsymbol{\phi}}\\\hat{\boldsymbol{\Sigma}}_{\boldsymbol{\theta}\boldsymbol{\phi}}^T & \hat{\boldsymbol{\Sigma}}_{\boldsymbol{\phi}\boldsymbol{\phi}}\end{bmatrix}\right) \qquad (19)$$

where $(\hat{\boldsymbol{\theta}},\hat{\boldsymbol{\phi}})$ denotes the MPV of the parameters (mode of the distribution), calculated by minimizing the logarithm of the posterior distribution, i.e., $L(\boldsymbol{\theta},\boldsymbol{\phi}) = -\ln p(\boldsymbol{\theta},\boldsymbol{\phi}|\mathbf{Y},\mathbf{X},\mathbb{M}_\mathbb{P})$; the matrix blocks $\hat{\boldsymbol{\Sigma}}_{\boldsymbol{\theta}\boldsymbol{\theta}} \in \mathbb{R}^{N_\theta \times N_\theta}$, $\hat{\boldsymbol{\Sigma}}_{\boldsymbol{\theta}\boldsymbol{\phi}} \in \mathbb{R}^{N_\theta \times N_\phi}$, and $\hat{\boldsymbol{\Sigma}}_{\boldsymbol{\phi}\boldsymbol{\phi}} \in \mathbb{R}^{N_\phi \times N_\phi}$ are calculated by inverting the Hessian matrix of $L(\boldsymbol{\theta},\boldsymbol{\phi})$ at the MPV. The approximation in Eq. (19) requires simultaneously optimizing both the structural parameters ($\boldsymbol{\theta}$) and the kernel function parameters ($\boldsymbol{\phi}$) over the augmented multi-dimensional space of these parameters. However, the simultaneous minimization of $\boldsymbol{\theta}$ and $\boldsymbol{\phi}$ makes the optimization likely to suffer from premature termination due to sticking into undesirable local minima.

As a remedy to this problem, we propose to break the minimization into a two-stage process, as provided in Algorithm 1, where $p(\boldsymbol{\theta}|\boldsymbol{\phi},\mathbf{Y},\mathbf{X},\mathbb{M}_\mathbb{P})$ and $p(\boldsymbol{\phi}|\boldsymbol{\theta},\mathbf{Y},\mathbf{X},\mathbb{M}_\mathbb{P})$ are maximized iteratively until convergence is attained. To find the mathematical expression of these probability distributions, the Bayes' rule can be used as follows:

$$p(\boldsymbol{\theta}|\boldsymbol{\phi},\mathbf{Y},\mathbf{X},\mathbb{M}_\mathbb{P}) = \frac{p(\mathbf{Y}|\boldsymbol{\theta},\boldsymbol{\phi},\mathbf{X},\mathbb{M}_\mathbb{P})p(\boldsymbol{\theta}|\boldsymbol{\phi},\mathbb{M}_\mathbb{P})}{p(\mathbf{Y}|\mathbf{X},\boldsymbol{\phi},\mathbb{M}_\mathbb{P})} \qquad (20)$$

Based on this equation, the logarithm of $p(\boldsymbol{\theta}|\boldsymbol{\phi},\mathbf{Y},\mathbf{X},\mathbb{M}_\mathbb{P})$ can be calculated as

$$\ln p(\boldsymbol{\theta}|\boldsymbol{\phi},\mathbf{Y},\mathbf{X},\mathbb{M}_\mathbb{P}) = \ln p(\mathbf{Y}|\boldsymbol{\theta},\boldsymbol{\phi},\mathbf{X},\mathbb{M}_\mathbb{P}) + \ln p(\boldsymbol{\theta}|\boldsymbol{\phi},\mathbb{M}_\mathbb{P}) - \ln p(\mathbf{Y}|\mathbf{X},\boldsymbol{\phi},\mathbb{M}_\mathbb{P}) \qquad (21)$$

Likewise, the probability distribution $p(\boldsymbol{\phi}|\boldsymbol{\theta},\mathbf{Y},\mathbf{X},\mathbb{M}_\mathbb{P})$ can be specified by exchanging the position of $\boldsymbol{\theta}$ and $\boldsymbol{\phi}$. Then, maximizing these conditional distributions turns out to be equivalent to minimizing the negative logarithm of these functions, as given by

$$L(\boldsymbol{\theta}|\boldsymbol{\phi}) \doteq -\ln p(\boldsymbol{\theta}|\boldsymbol{\phi},\mathbf{Y},\mathbf{X},\mathbb{M}_\mathbb{P}) = \frac{1}{2}(\mathbf{Y}-\mathbf{f}(\mathbf{X};\boldsymbol{\theta}))^T\mathbf{K}^{-1}(\mathbf{Y}-\mathbf{f}(\mathbf{X};\boldsymbol{\theta})) - \ln p(\boldsymbol{\theta}|\boldsymbol{\phi},\mathbb{M}_\mathbb{P}) + cte. \qquad (22)$$

$$L(\boldsymbol{\phi}|\boldsymbol{\theta}) \doteq -\ln p(\boldsymbol{\phi}|\boldsymbol{\theta},\mathbf{Y},\mathbf{X},\mathbb{M}_\mathbb{P}) = \frac{1}{2}\ln|\mathbf{K}| + \frac{1}{2}(\mathbf{Y}-\mathbf{f}(\mathbf{X};\boldsymbol{\theta}))^T\mathbf{K}^{-1}(\mathbf{Y}-\mathbf{f}(\mathbf{X};\boldsymbol{\theta})) - \ln p(\boldsymbol{\phi}|\boldsymbol{\theta},\mathbb{M}_\mathbb{P}) + cte. \qquad (23)$$

where *cte.* stands for constant terms; the determinant term in the Eq. (22) was dropped since it depends only on the parameters ($\boldsymbol{\phi}$) and does not affect the optimization process when calculating the parameters ($\boldsymbol{\theta}$).

Algorithm 1 formalizes the process of finding the MPV of the parameters, which reads as a main "while-loop" iteratively updating ($\boldsymbol{\theta}$) and ($\boldsymbol{\phi}$). In this algorithm, the convergence is reached when the variation of the estimated parameters between the last two iterations of the algorithm is no longer above a predefined threshold. This condition should be checked along with a stopper placed on the total number of iterations, aiming to avoid too many trials. Once the MPV of the parameters are obtained, we can predict structural responses by replacing the optimal values into Eqs. (7-10) while ignoring their uncertainties. Algorithm 2 provides the detailed steps of predicting responses based on the inferred parameters. Although this procedure would be inaccurate due to neglecting the uncertainty of the parameters, it offers significant computational savings. In case a more rigorous calculation is required, a full sampling approach might be preferred, as explained in the next section.

**Algorithm 1.** Calculation of the MPV of the parameters
1. Set initial values of the structural parameters ($\boldsymbol{\theta}^{(0)}$) and GP model ($\boldsymbol{\phi}^{(0)}$).
2. Set the convergence measure (*Conv*) to one.
3. Set the maximum number of iterations, e.g., $Max = 100$.
4. Set the iteration number of one ($j = 0$).
5. Set the convergence tolerance to a small number, e.g., $Tol = 10^{-6}$
6. **While** ($Conv > Tol$ & $j < Max$) {
7. Calculate the inverse of the covariance matrix ($\hat{\mathbf{K}}^{-1}$) based on the latest estimate of ($\boldsymbol{\phi}^{(j)}$).
8. Minimize $L(\boldsymbol{\theta} | \boldsymbol{\phi}^{(j)})$ given by Eq. (22) and calculate the optimal values $\boldsymbol{\theta}^{(j+1)} = \text{Argmin} \, L(\boldsymbol{\theta} | \boldsymbol{\phi}^{(j)})$.
9. Minimize $L(\boldsymbol{\phi} | \boldsymbol{\theta}^{(j+1)})$ given by Eq. (23) and calculate the optimal values $\boldsymbol{\phi}^{(j+1)} = \text{Argmin} \, L(\boldsymbol{\phi} | \boldsymbol{\theta}^{(j+1)})$.
10. Calculate $L_1 = L(\boldsymbol{\theta}, \boldsymbol{\phi}) = -\ln p(\boldsymbol{\theta}, \boldsymbol{\phi} | \mathbf{Y}, \mathbf{X}, \mathbb{M}_\mathbb{P})$ using the updated estimations, $\boldsymbol{\theta}^{(j+1)}$ and $\boldsymbol{\phi}^{(j+1)}$.
11. Calculate $Conv = \| [\boldsymbol{\theta}^{(j+1),T} \, \boldsymbol{\phi}^{(j+1),T}]^T - [\boldsymbol{\theta}^{(j),T} \, \boldsymbol{\phi}^{(j),T}]^T \| / \| [\boldsymbol{\theta}^{(j),T} \, \boldsymbol{\phi}^{(j),T}]^T \|$
12. Set $j = j+1$.
13. } **End while**

**Algorithm 2.** Response predictions using MAP estimations
1. Provide $\hat{\boldsymbol{\theta}}$ and $\hat{\boldsymbol{\phi}}$ from Algorithm 1.
2. Calculate $\hat{\mathbf{K}}_{pred}$ and $\hat{\mathbf{k}}_{pred}$ by using predefined kernel covariance function.
3. Calculate the predicted mean $\hat{\boldsymbol{\mu}}_{pred} = \mathbf{f}(\mathbf{X}_{pred}; \hat{\boldsymbol{\theta}}) + \hat{\mathbf{k}}_{pred}^T \hat{\mathbf{K}}_{pred}^{-1}(\mathbf{Y} - \mathbf{f}(\mathbf{X}; \hat{\boldsymbol{\theta}}))$.
4. Calculate the covariance matrix $\hat{\boldsymbol{\Sigma}}_{pred} = \hat{\mathbf{K}}_{pred} - \hat{\mathbf{k}}_{pred}^T \hat{\mathbf{K}}^{-1} \hat{\mathbf{k}}_{pred}$
5. Provide the posterior predictive distribution $p(\mathbf{Y}_{pred} | \mathbf{Y}, \mathbf{X}, \mathbf{X}_{pred}, \mathbb{M}_\mathbb{P}) \approx N(\mathbf{Y}_{pred} | \hat{\boldsymbol{\mu}}_{pred}, \hat{\boldsymbol{\Sigma}}_{pred})$.

## 3.2. Full Sampling Strategy

MCMC sampling methods can be regarded as a more general treatment of the inference problem formulated above. In this study, we use the Transitional Markov Chain Monte Carlo (TMCMC) method [58–60] for drawing samples from the posterior distribution described by Eqs. (3-5). By doing so, the mean and covariance of the unknown parameters can be estimated using the statistics of the posterior samples:

$$\mathbb{E}\left(\begin{bmatrix}\boldsymbol{\theta}\\\boldsymbol{\phi}\end{bmatrix}\right) \approx \frac{1}{N_s}\sum_{m=1}^{N_s}\begin{bmatrix}\boldsymbol{\theta}^{(m)}\\\boldsymbol{\phi}^{(m)}\end{bmatrix} \quad (24)$$

$$\mathbb{COV}\left(\begin{bmatrix}\boldsymbol{\theta}\\\boldsymbol{\phi}\end{bmatrix}\right) \approx \frac{1}{N_s}\sum_{m=1}^{N_s}\left(\begin{bmatrix}\boldsymbol{\theta}^{(m)}\\\boldsymbol{\phi}^{(m)}\end{bmatrix} - \mathbb{E}\left(\begin{bmatrix}\boldsymbol{\theta}\\\boldsymbol{\phi}\end{bmatrix}\right)\right)\left(\begin{bmatrix}\boldsymbol{\theta}^{(m)}\\\boldsymbol{\phi}^{(m)}\end{bmatrix} - \mathbb{E}\left(\begin{bmatrix}\boldsymbol{\theta}\\\boldsymbol{\phi}\end{bmatrix}\right)\right)^T \quad (25)$$

where $\mathbb{E}[.]$ and $\mathbb{COV}[.]$ denote the expected value (mean) and the covariance matrix, respectively; $\boldsymbol{\theta}^{(m)}$ and $\boldsymbol{\phi}^{(m)}$ are the samples obtained from the posterior distribution $p(\boldsymbol{\theta}, \boldsymbol{\phi} | \mathbf{Y}, \mathbf{X}, \mathbb{M}_\mathbb{P})$. Additionally, the integrals encountered for calculating the evidence term in Eq. (4) can be approximated as

$$p(\mathbf{Y} | \mathbf{X}, \mathbb{M}_\mathbb{P}) \approx \frac{1}{N_s}\sum_{m=1}^{N_s} p(\mathbf{Y} | \boldsymbol{\theta}^{(m)}, \boldsymbol{\phi}^{(m)}, \mathbf{X}, \mathbb{M}_\mathbb{P}) \quad (26)$$

Although this expression is generally true, calculation of the evidence term based on this approximation might not be sufficiently accurate, especially when the likelihood function is highly peaked [58–60]. Therefore, in this paper, we resort to the TMCMC sampling algorithms [58–60] to calculate it in a more reliable manner. In this approach, in lieu of attempting to converge from a prior distribution directly into a target PDF, a series of intermediate proposal distributions are created to draw samples from the target distribution iteratively. The intermediate distributions are determined through an additional resampling stage, where the samples are synthetized based on plausibility weights. In most cases, the intermediate distributions are considered Gaussian centered at each sample of the Markov chain, whose spread is characterized via a scaled covariance matrix. Further details about TMCMC sampler and its technical details are available elsewhere, e.g., [58–60].

Once the evidence term is calculated, it is straightforward to use Eq. (11) for performing Bayesian model class selection. Furthermore, the dynamical response of the structure subjected to new input scenarios can be approximated based on Eqs. (7-10) and the sampling results, giving:

$$p(\mathbf{Y}_{pred} \mid \mathbf{Y}, \mathbf{X}, \mathbf{X}_{pred}, \mathbb{M}_{\mathbb{P}}) \approx \frac{1}{N_s} \sum_{s=1}^{N_s} N\left(\mathbf{Y}_{pred} \mid \boldsymbol{\mu}_{pred}^{(m)}, \boldsymbol{\Sigma}_{pred}^{(m)}\right) \tag{27}$$

where $\boldsymbol{\mu}_{pred}^{(m)} \in \mathbb{R}^{n'N_o}$ and $\boldsymbol{\Sigma}_{pred}^{(m)} \in \mathbb{R}^{n'N_o \times n'N_o}$ are respectively the mean vector and covariance matrix of the response calculated based individual samples $\boldsymbol{\theta}^{(m)}$ and $\boldsymbol{\phi}^{(m)}$, given by

$$\boldsymbol{\mu}_{pred}^{(m)} = \mathbf{f}(\mathbf{X}_{pred}; \boldsymbol{\theta}_{pred}^{(m)}) + (\mathbf{k}_{pred}^{(m)})^T (\mathbf{K}_{pred}^{(m)})^{-1} (\mathbf{Y} - \mathbf{f}(\mathbf{X}; \boldsymbol{\theta}_{pred}^{(m)})) \tag{28}$$

$$\boldsymbol{\Sigma}_{pred}^{(m)} = \mathbf{K}_{pred}^{(m)} - (\mathbf{k}_{pred}^{(m)})^T (\mathbf{K}_{pred}^{(m)})^{-1} \mathbf{k}_{pred}^{(m)} \tag{29}$$

It simply follows from Eq. (29) that the mean and covariance of the Gaussian mixture distribution can be calculated explicitly [32]:

$$\mathbb{E}\left(\left[\mathbf{Y}_{pred}\right]\right) = \frac{1}{N_s} \sum_{s=1}^{N_s} \boldsymbol{\mu}_{pred}^{(m)} \tag{30}$$

$$\mathbb{COV}\left(\left[\mathbf{Y}_{pred}\right]\right) = \frac{1}{N_s} \sum_{s=1}^{N_s} \left((\boldsymbol{\mu}_{pred}^{(m)})(\boldsymbol{\mu}_{pred}^{(m)})^T + \boldsymbol{\Sigma}_{pred}^{(m)}\right) - \mathbb{E}\left[\mathbf{Y}_{pred}\right]\mathbb{E}\left[\mathbf{Y}_{pred}^T\right] \tag{31}$$

Algorithm 3 puts together the flow steps for model inference and response predictions using MCMC sampling methods. In this algorithm, as expected, the computational cost is mainly concerned with step "1" of the algorithm, where the MCMC samples are obtained. The remaining steps are mainly post-processing, which end up with the estimated second-moment statistics of the parameters and responses.

**Algorithm 3.** MCMC sampling approach
1. Draw samples $(\boldsymbol{\theta}^{(m)}, \boldsymbol{\phi}^{(m)})$ from posterior distribution $p(\boldsymbol{\theta}, \boldsymbol{\phi} \mid \mathbf{Y}, \mathbf{X}, \mathbb{M}_{\mathbb{P}})$, where $m = \{1, ..., N_s\}$.
2. Estimate the mean and covariance of the parameters based on Eq. (24-25).
3. **For** $m = 1 : N_s$, do the following:
4. Calculate $\mathbf{k}_{pred}^{(m)}$ and $\mathbf{K}_{pred}^{(m)}$ for each sample $\boldsymbol{\phi}^{(m)}$ using the predefined kernel covariance function.
5. Calculate samples of the predictive mean and covariance of the response from Eqs. (28-29).
6. **End For**
7. Calculate second-moment statistics of the response using Eqs. (30-31).

### 3.3. Discussion on Computational Costs
The above optimization and sampling algorithms involve a large number of evaluation of the log-likelihood function. This process is computationally expensive since it needs running the structural model to produce dynamical response and calculating the determinant and inverse of the covariance matrix. The former can be alleviated through model reduction techniques [61] and surrogate modeling techniques (e.g., [62]). The latter can easily overwhelm computations even in low-dimensional problems as it scales cubically with the number of data points, i.e., $O(n^3 N_o^3)$, leading to excessively high dimensional matrices when dealing with long time-history responses. In the machine learning literature, it is customary to use Cholesky decomposition of the covariance matrices and calculate the determinant and inverse matrix [41]. However, this decomposition can still be computationally prohibitive, especially when dealing with time-domain vibration data. Moreover, in this approach, the optimization becomes very sensitive to the initial values, and the estimated parameters might vary significantly. This issue might occur due to large noise contamination of the signal, which usually prevails in the measurements. Therefore, in this work, the Singular Value Decomposition (SVD) of the covariance matrix is employed for truncating redundant dimensions from the observed data covariance matrix and alleviate scalability problems. In order to specify an appropriate level of truncation, the slope of the singular values decay (gradient) is checked against a tolerance of 0.1-1% of the largest singular values. Later, we will illustrate how such a dimensionality reduction approach can be used.

## 4. Numerical Example I

The acceleration response of a linear single-degree-of-freedom (SDOF) system is generated synthetically subjected to a known input force. The input force is zero-mean GWN with 1N standard deviation shown in Fig. 2(a). The mass is 1kg, the stiffness is 5N/m, and the viscous damping ratio is 5%. The sampling interval is considered 0.01s, and the response length is 40s shown in Fig. 2(b).

The structural model considered is a linear SDOF structure, having 1kg mass and $\theta$ N/m stiffness, where $\theta$ should be identified. The viscous damping ratio is set at 4%, considered unequal to the damping of the structure. This misspecification aims to introduce mismatch between the model and measured responses for all choices of $\theta$. Thus, there is no correct value for the parameter ($\theta$) is to be identified, but an estimate close to 5N/m is desirable. For $\theta = 5$N/m, the model response is very close to the presumed measured response, as shown in Fig. 2(b). The discrepancy between the generated and model responses (prediction error process) is also shown in Fig. 2(c), which seems to be a sinusoidal function with a dominant frequency.

Given the above assumptions, four types of kernel functions, GWN, SE, PE, and MMTE are considered for model inference and response prediction in this example. Note that, when using SE, PE, and MMTE, they are composed with an additive GWN to capture residual random effects. Moreover, the auxiliary parameters ($\zeta_i$) is considered to be the same as time steps. As the SDOF system has only one mode in its responses, in the MMT kernel function, the model order is set to one, i.e., $m = 1$. The first 20s of the data is used for model inference as training data, and the other 20s segment of the data is used for demonstrating prediction properties of the kernel functions. The results of applying Algorithms 1-3 are summarized in Table 1. The MPV ($\hat{\theta}$) calculated by optimization and the estimated mean $\mathbb{E}[\theta]$ identified from iTMCMC sampling [59] are very close to the actual value (5N/m), regardless of the choice of the covariance function. However, the uncertainty of the stiffness, represented by $\hat{\sigma}_\theta$ (inverse of the Hessian matrix) or $\text{SD}[\theta]$ (standard deviation of the samples), differs depending on the choice of the kernel function.

The logarithm of the evidence term and the posterior probability is calculated and reported in Table 1 for each kernel function by using Laplace approximation and sampling approaches. According to this table, the MMTE is the preferred case. This result agrees with response prediction shown in Fig. 3(a-d), where the MMTE outperforms other kernel functions. As shown, the GWN kernel function gives zero mean error and a fixed uncertainty bound. The SE initially provides good accuracy, but due to characterizing a small correlation length, it loses accuracy and gives zero mean predictions along with a large uncertainty bound. The PE repeats the same signal as the training data, falling off the actual prediction error process; however, the MMTE well captures the periodic pattern in the prediction error process. The uncertainty given by the MMTE is initially small but increases as time goes by and as the accuracy of the estimated mean error deteriorates.

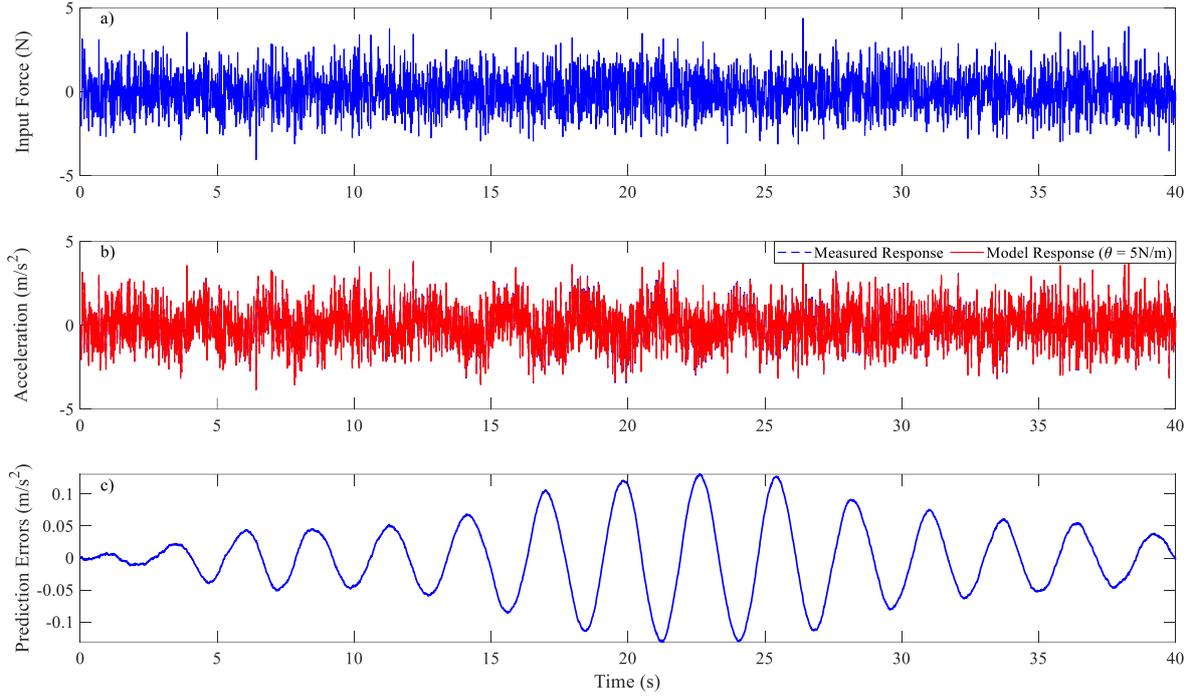

**Fig. 2.** (a) Input force (b) Measured and model responses (c) Discrepancy between measured and model responses

**Table 1.**
Identification results for different choices of kernel covariance function

| Method | Calculated Quantities | Kernel Covariance Function | | | |
|---|---|---|---|---|---|
| | | GWN | SE | PE | MMTE |
| Laplace Approximation | $\hat{\theta}$ (MPV) | 4.993 | 5.006 | 5.009 | 5.001 |
| | $\hat{\sigma}_\theta$ (SD) | 0.0017 | 0.005 | 0.004 | 0.014 |
| | $\hat{\ell}$ | - | 0.391 | 5.842 | 0.189 |
| | $\hat{\sigma}_f$ | - | 0.060 | 0.091 | 0.059 |
| | $\hat{\omega}$ | - | - | 8.2943 | 2.255 |
| | $\hat{\sigma}_n$ | 0.0467 | $8.5 \times 10^{-7}$ | $9.9 \times 10^{-5}$ | $1.9 \times 10^{-9}$ |
| | $\ln p(\mathbf{Y} \mid \mathbf{X}, \mathbb{M}_\mathbb{P})$ | $5.1 \times 10^3$ | $8.4 \times 10^3$ | $-8.6 \times 10^3$ | $1.3 \times 10^4$ |
| | $\ln P(\mathbb{M}_\mathbb{P} \mid \mathbf{X}, \mathbf{X}_{pred}, \mathbf{Y}, \mathbf{Y}_{pred})$ | $9.3 \times 10^3$ | $1.7 \times 10^4$ | $-1.2 \times 10^{15}$ | $2.7 \times 10^4$ |
| TMCMC Sampling | $\mathbb{E}[\theta]$ | 5.003 | 5.017 | 4.946 | 5.000 |
| | $SD[\theta]$ | 0.0004 | 0.0194 | 0.0175 | 0.0220 |
| | $\ln p(\mathbf{Y} \mid \mathbf{X}, \mathbb{M}_\mathbb{P})$ | $-3.3 \times 10^4$ | $3.9 \times 10^3$ | $-4.8 \times 10^3$ | $1.3 \times 10^4$ |
| | $\ln P(\mathbb{M}_\mathbb{P} \mid \mathbf{X}, \mathbf{X}_{pred}, \mathbf{Y}, \mathbf{Y}_{pred})$ | $-2.9 \times 10^4$ | $1.3 \times 10^4$ | $-1.8 \times 10^{14}$ | $2.5 \times 10^4$ |

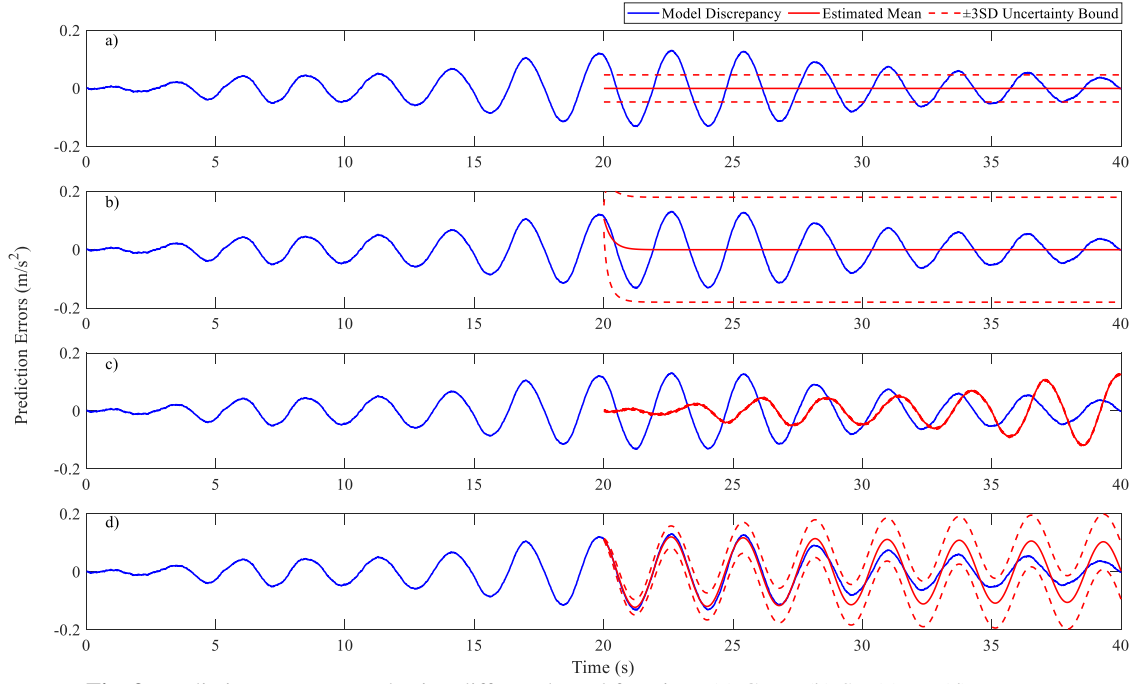
**Fig. 3.** Prediction errors created using different kernel functions (a) GWN (b) SE (c) PE (d) MMTE

The reason for seeing that the MMTE is superior to other kernels should be searched in the temporal characteristics of the prediction errors. For this purpose, the prediction error is calculated by considering $\theta = 5$ N/m, and then, the sample autocorrelation function and power spectral density (PSD) are calculated. As shown in Fig. 4(a-b), the PSD of the prediction error has a sharp peak very close to the natural frequency of the structure $\sqrt{5}/(2\pi)$ Hz, exhibiting that a sinusoidal kernel function like MMTE suits the problem at hand.

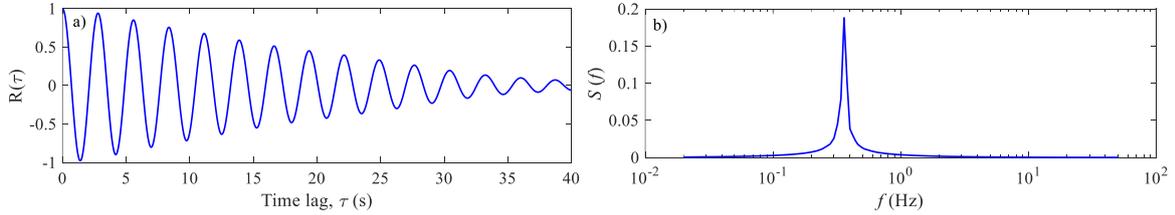
**Fig. 4.** (a) Sample autocorrelation function of the prediction error (b) Sample PSD.

As suggested in Section 3.3, the covariance matrix can be reduced by removing redundant dimensions. Fig. 5 shows the eigenvalues of the covariance matrix ($\mathbf{K}$) on a logarithmic scale, displaying a fast decay in the magnitudes. The cut-off threshold prescribed earlier implies keeping 30 singular values. This threshold can be a good choice since the truncated eigenvalues are smaller than the largest one up to several orders of magnitude.

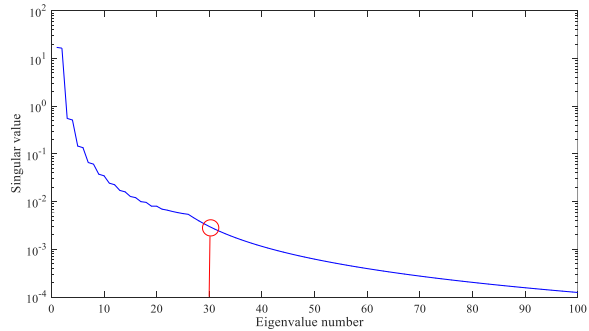
**Fig. 5.** Eigenvalues of the covariance matrix and the truncation threshold

## 4.1. Effects of Larger Modeling Error

In the example presented earlier, the modeling error was relatively small, created by considering 4% damping ratio while the system was 5% damped. In this section, we display the results for other choices of damping ratios while using the same set of assumptions as above. Fig. 6(a-c) demonstrates model discrepancy considering 1%, 2%, and 3% damping ratios, respectively. It is evident that the predictions are properly accurate, and the uncertainty bounds consistently account for the errors, regardless of the value of damping ratios assumed.

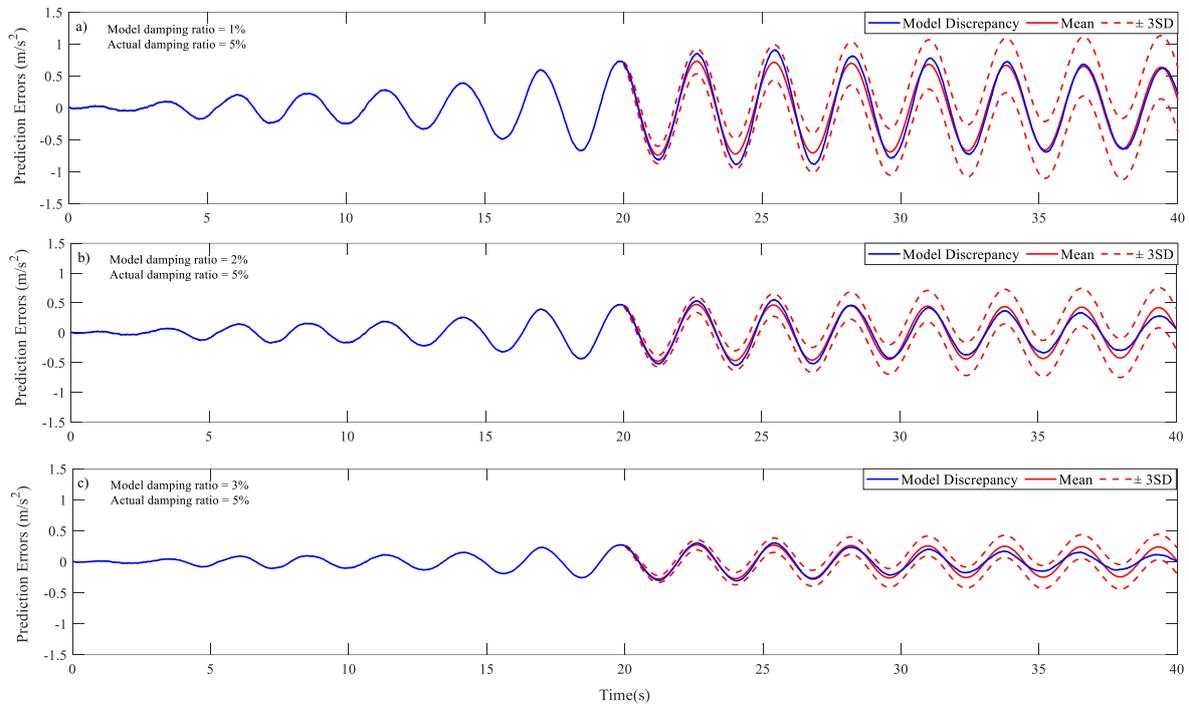

**Fig. 6.** Performance comparison of the proposed method when using 1%, 2%, and 3% damping ratios for the structural model class

## 4.2. Missing Data Samples

An important aspect of the proposed framework is the reconstruction of missing data based on observed segments of data. This issue might happen in practice due to malfunction of sensing or data acquisition devices. In this example, we have assumed that two segments of data [0-20s] and [30-40s] are available while the segment [20-30s] is missing. The MMTE-based GP model is used for estimating the missing segment. Fig. 7(a) shows high accuracy of the reconstructed segment. The interval [20-22s] is magnified in Fig. 7(b), where the mean and ±SD uncertainty bounds are included to indicate the accuracy of the predictions. In Fig. 7(c), the prediction error process is plotted along with the mean error and uncertainty bounds. As can be seen, the proposed framework accurately estimates the missing data in the presence of modeling errors.

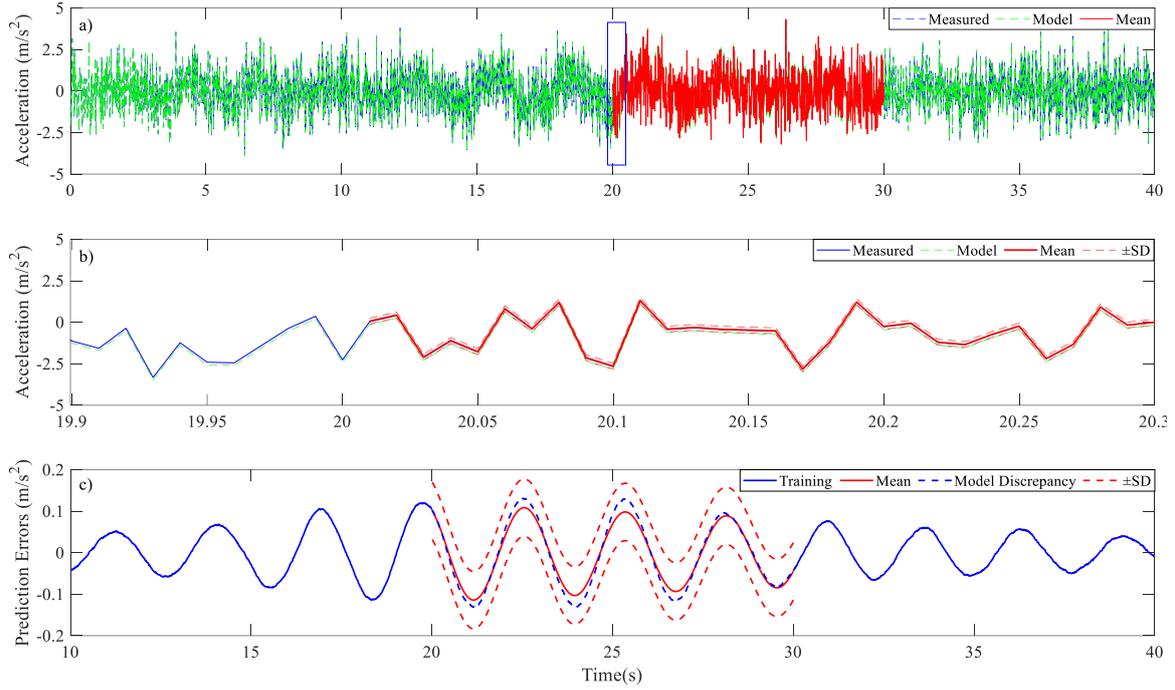

**Fig. 7.** Reconstruction of missing acceleration response history

## 5. Numerical Example II
### 5.1. Basic Assumptions

In this section, a five-story structure shown in the Fig. 8 is used for testing the proposed framework. Since horizontal beams act rigidly, the only DOFs are the lateral displacements of the floors shown in Fig. 8. The mass matrix is diagonal with equal elements of 1kg. The stiffness of all stories is considered $k_1 = ... = k_5 = 10\text{N/m}$. Thus, the modal frequencies of the structure can be calculated as $\Omega = \{0.90, 2.63, 4.14, 5.32, 6.07\}$ rad/s. The damping matrix is proportional to the mass and stiffness matrices, i.e., $\mathbf{C} = \alpha\mathbf{K} + \beta\mathbf{M}$, where $\alpha = 0.02$ and $\beta = 2 \times 10^{-5}$. The structure is subjected to a dynamic force applied to the top floor, and the dynamical responses were simulated subjected to a GWN force with zero mean and 1N standard deviation using $\Delta t = 0.01s$ interval and $T = 60s$ duration. Two sensor configurations are considered:
- Configuration (A): one accelerometer placed on the 2$^{nd}$ story
- Configuration (B): two accelerometers placed on the 2$^{nd}$ and 4$^{th}$ stories

A structural model with a similar configuration is considered for performing model updating and response predictions. The mass matrix is known, and the structural model deviates from the actual structure in using a different damping coefficient of $\alpha' = 0.03$. The stiffness matrix is partially known, and the following cases are considered for illustrating the proposed method:
- Case (I): Only the stiffness of the 1$^{st}$ floor is unknown
- Case (II): the stiffness of the 1$^{st}$ and 4$^{th}$ floors are unknown
- Case (III): the stiffness of the 1$^{st}$, 4$^{th}$, and 5$^{th}$ floors are unknown

Thus, the objective is to update the stiffness parameters and predict the future response of the system in the presence of modeling errors while using measured input-output data. For simplicity, the unknown stiffness is replaced by their nominal values multiplied by the dimension-less parameters $\boldsymbol{\theta}$, which should be identified. Thus, the unknown parameters ($\boldsymbol{\theta}$) are expected to be identified very close to one.

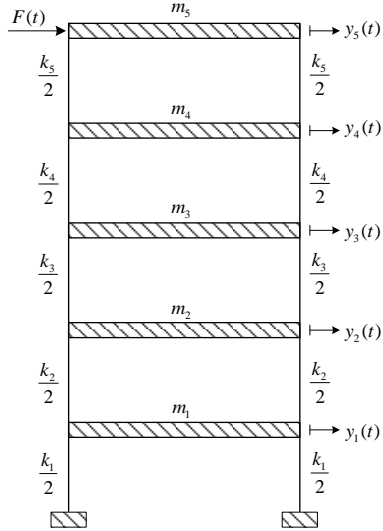

**Fig. 8.** Shear frame used for illustrating the proposed methodology

### 5.2. Results: Case (I) using Sensor Configuration (A)

We first apply the proposed framework to Case (I) while using sensor configuration (A). Having observed the success of the MMTE kernel function in the preceding section, we only demonstrate the GP framework using this kernel function. This choice can be justified by taking a careful look at the sample autocorrelation function of the prediction errors. As shown in the Fig. 9(a), the autocorrelation constitutes a few mixing sinusoidal functions. This pattern can better be seen in Fig. 9(b), where the Fourier transform of the autocorrelation function is indicated, showing the PSD of the prediction errors. It can be confirmed that there are four dominant peaks, each potentially representing a separate cosine function. Although it might deem reasonable to consider the order of the kernel function to be $m=4$, the complexity of the kernel covariance function requires additional investigation in order to avoid over-fitting due to extra parameterization of the kernel function.

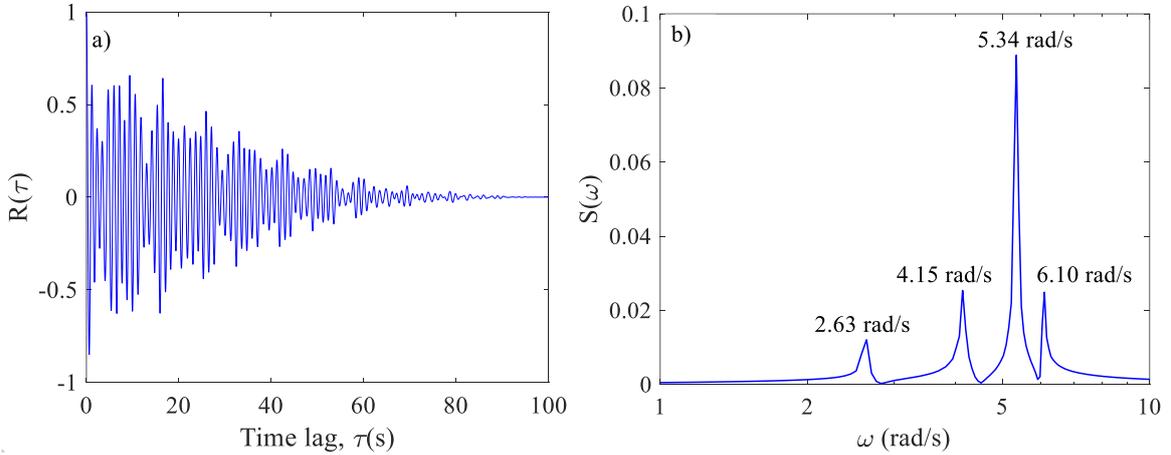

**Fig. 9.** (a) Sample Auto-correlation function (ACF) of the prediction error (b) PSD of the prediction errors

At this point, we clarify that the order of the MMTE kernel characterizes the number of dominant peaks on the PSD curves. Thus, considering a large order for this covariance function can be interpreted as modeling more peaks than necessary. In this case, it is likely that spurious peaks appearing due to measurement noise be realized as dominant periodic patterns, making the quality of predictions inferior to the case wherein a smaller model order is selected. Additionally, since the number of unknown parameters also grows with the order of the kernel function, a large order creates extra computational costs that might be unnecessary.

To choose an optimal model order for the kernel function, a Bayesian model selection perspective is adopted herein. For this purpose, the evidence term is approximated through the Bayesian Information Criteria (BIC) using the following formula:

$$P(\mathbb{M}_\mathbb{P} | \mathbf{X}, \mathbf{Y}) \approx BIC(\mathbb{M}_\mathbb{P}) = \ln p(\mathbf{Y} | \hat{\boldsymbol{\theta}}, \hat{\boldsymbol{\phi}}, \mathbf{X}, \mathbb{M}_\mathbb{P}) - \frac{1}{2}(N_\theta + N_\phi + 1)\ln(nN_o) \quad (30)$$

where $\ln p(\mathbf{Y} | \hat{\boldsymbol{\theta}}, \hat{\boldsymbol{\phi}}, \mathbf{X}, \mathbb{M}_\mathbb{P})$ is the likelihood function to which the MPV $\hat{\boldsymbol{\theta}}$ and $\hat{\boldsymbol{\phi}}$ are replaced, representing a measure-of-fit; the expression $\frac{1}{2}(N_\theta + N_\phi + 1)\ln(nN_o)$ penalizes parameterization, growing with the number of data and model parameters. Fig. 10 shows the *BIC* scores, as well as the log-evidence obtained using TMCMC versus the number of modes contributing to the covariance function. As can be seen, both methods confirm that $m = 3$ is a reasonable choice.

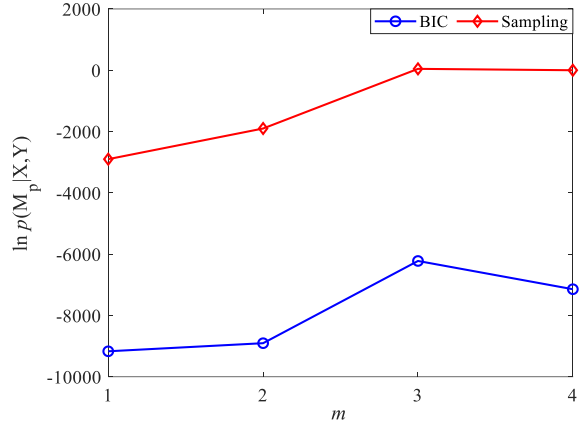

**Fig. 10.** Selection of the number of contributing modes into the kernel function

Since a similar trend is observed for other case studies and sensor configurations, $m = 3$ is considered for all cases. Thus, $\boldsymbol{\phi} = [\sigma_{f,1}^2 \; \omega_1 \; \ell_1^{-2} \; \sigma_{f,2}^2 \; \omega_2 \; \ell_2^{-2} \; \sigma_{f,3}^2 \; \omega_3 \; \ell_3^{-2} \; \sigma_n^2]^T$ should be calibrated based on the data. For this purpose, the iTMCMC sampling technique [59] was used to draw samples from the posterior distribution using Algorithm 3. The prior distribution of the parameters is considered uniform, accepting equal plausibility of the parameters over a relatively large domain. In Fig. 11, the posterior samples of $\boldsymbol{\phi}$ are displayed for Case (I) of model inference using data from sensor configuration (A). The diagonal plots show the histograms of the parameters, and the lower diagonal plots show joint posterior distributions in a pairwise manner. Based on this figure, the posterior samples of the parameters $\{\omega_1, \omega_2, \omega_3\}$ concentrate close to those of the structure's modal frequencies reported above. The samples of $\{\ell_1^{-2}, \ell_2^{-2}, \ell_3^{-2}\}$ populate around 0.02, and this result can be interpreted as having correlation lengths of around 7.1s, indicating how far we can expect to have reasonable predictions.

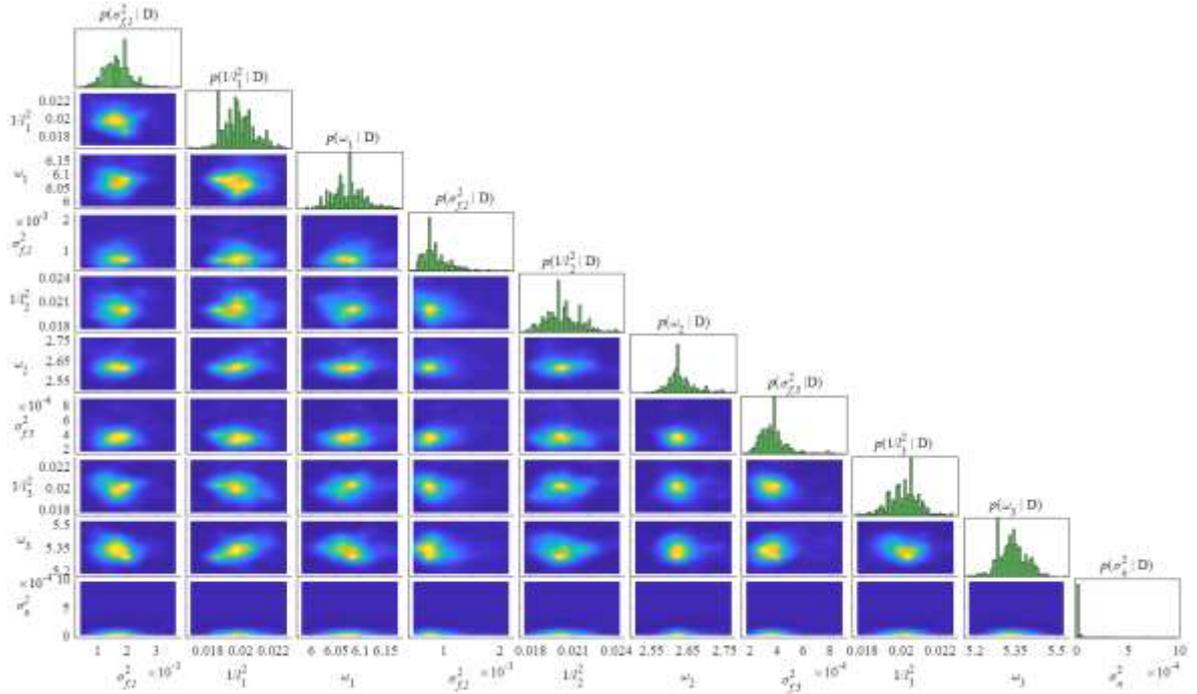

**Fig. 11.** Posterior samples of the hyper-parameters of the kernel function using $N_s = 5000$ number of samples.

## 5.3. Results: Comparison between Other Cases

Similar pattern was observed for other cases and configurations, but they are not provided in the paper for brevity. Table 2 presents the identification of stiffness parameters for different case studies, obtained using Algorithm 1. The MPV match the actual values, highlighting validity of the results. The posterior uncertainties are also calculated using Laplace approximation. It can be shown that the uncertainty in the parameters consistently increases with the number of unknown parameters while reduces when additional sensors are included.

**Table 2.**
Estimates of mean and standard deviation of the stiffness ratios using sampling for the signal trained at length of $T = 30s$ and $\Delta t = 0.01s$

| Case ID | $\hat{\theta}_1$ | $\hat{\sigma}_{\theta_1}$ | $\hat{\theta}_2$ | $\hat{\sigma}_{\theta_2}$ | $\hat{\theta}_3$ | $\hat{\sigma}_{\theta_3}$ |
|---|---|---|---|---|---|---|
| **Sensor Configuration (A)** | | | | | | |
| (I)   | 1.0020 | 0.0033 | -      | -      | -      | -      |
| (II)  | 1.0056 | 0.0042 | 1.0034 | 0.0043 | -      | -      |
| (III) | 1.0040 | 0.0195 | 0.9944 | 0.0113 | 1.0004 | 0.0250 |
| **Sensor Configuration (B)** | | | | | | |
| (I)   | 0.9995 | 0.0018 | -      | -      | -      | -      |
| (II)  | 1.0005 | 0.0025 | 0.9988 | 0.0025 | -      | -      |
| (III) | 1.0020 | 0.0039 | 0.9983 | 0.0034 | 0.9992 | 0.0048 |

Once the unknown parameters are inferred through Laplace approximation or sampling, it is possible to perform response predictions. Figs. 12(a) and 13(a) show the predicted resposes of the 2$^{nd}$ and 4$^{th}$ stories. Figs. 12(b) and 13(b) compare the mean and uncertainty bounds along with the model discrepancy. The accuracy and robustness of predictions are remarkablely good, and the periodic pattern in the discrepancy model is well captured via the proposed GP model. However, due to loss of correlation with the data, the accuracy of the estimated mean deteriorates over time. It seems the trusted length of predictions can be up to the same order as the correaltion length identified by the GP model.

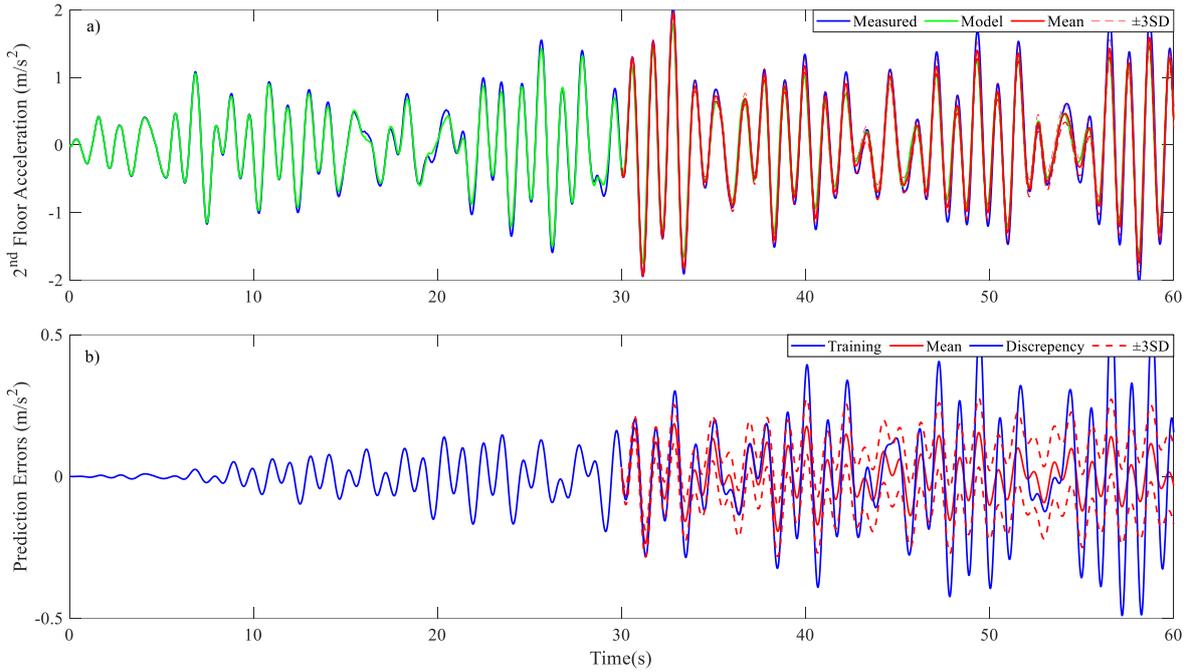

**Fig. 12.** Prediction of the response time histories and the errors of the Second floor (Training based on Case (III) and data from Configuration (B))

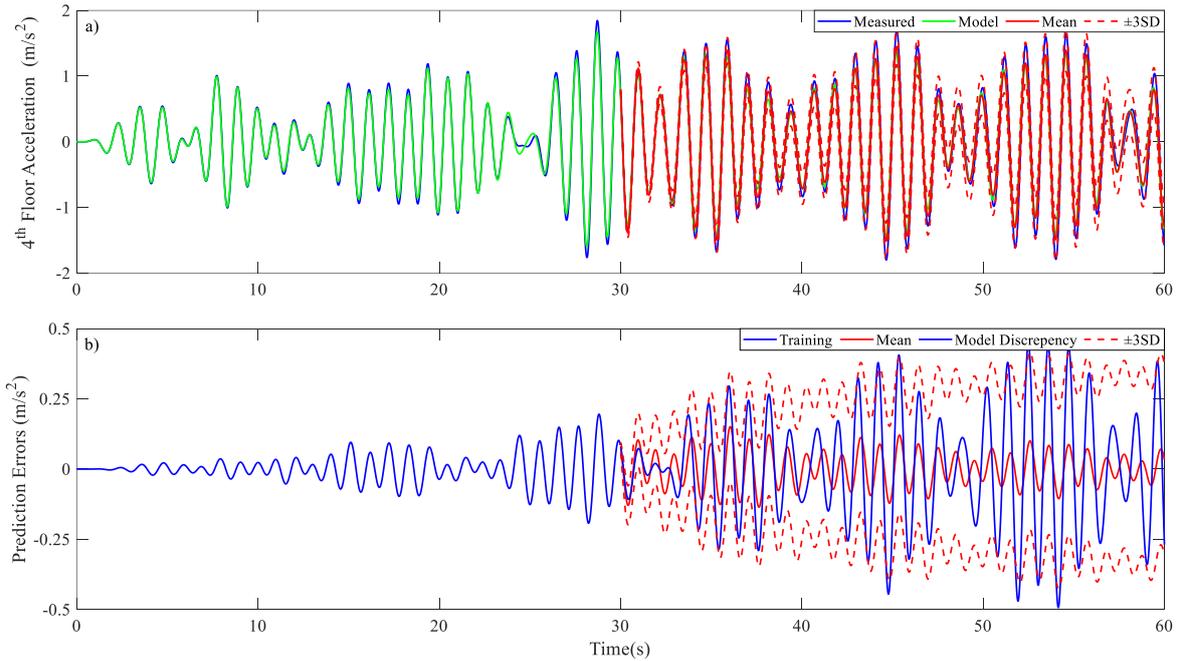

**Fig. 13.** Prediction of the response time histories and the errors of the Fourth floor (Training based on Case (III) and Configuration (B))

## 6. Experimental Example

In this section, the proposed framework is illustrated using experimental data from the three-story structure shown in Fig. 14(a). A shaking table test is performed to apply base excitation and record dynamical responses. The acceleration responses of the base and three stories were measured under a band-limited GWN base excitation. The data is sampled at 0.005s intervals and recorded for 120s. The mass of the structure is lumped at the floor levels, measured as 5.63kg, 6.03kg, and 4.66kg corresponding to the first, second, and third stories, respectively. The mass of other members is ignored as is considerably smaller than floor masses.

A shear frame model shown in Fig. 14(b) is used for describing the dynamics. The mass of the structural model is known as $\mathbf{M} = \text{diag}([5.63\ 6.03\ 4.66])$ while the stiffness and damping matrices have to be identified considering the following parameterization:

$$\mathbf{K} = \begin{bmatrix} k_1 + k_2 & -k_2 & 0 \\ -k_2 & k_2 + k_3 & -k_3 \\ 0 & -k_3 & k_3 \end{bmatrix} \quad (31)$$

$$\mathbf{C} = \sum_{i=1}^{3} 2\zeta_i \hat{\omega}_i \frac{\mathbf{M}\hat{\boldsymbol{\varphi}}_i \hat{\boldsymbol{\varphi}}_i^T \mathbf{M}}{\hat{\boldsymbol{\varphi}}_i^T \mathbf{M} \hat{\boldsymbol{\varphi}}_i} \quad (32)$$

where $\hat{\omega}_i$ and $\hat{\boldsymbol{\varphi}}_i$ are nominal values of the modal frequency and mode shape corresponding to the $i^{\text{th}}$ dynamical mode, obtained from [63]; $\zeta_i$ is the $i^{\text{th}}$ unknown modal damping ratio. Given this specification, the unknown parameters can be collected into $\boldsymbol{\theta} = [k_1\ k_2\ k_3\ \zeta_1\ \zeta_2\ \zeta_3]^T$.

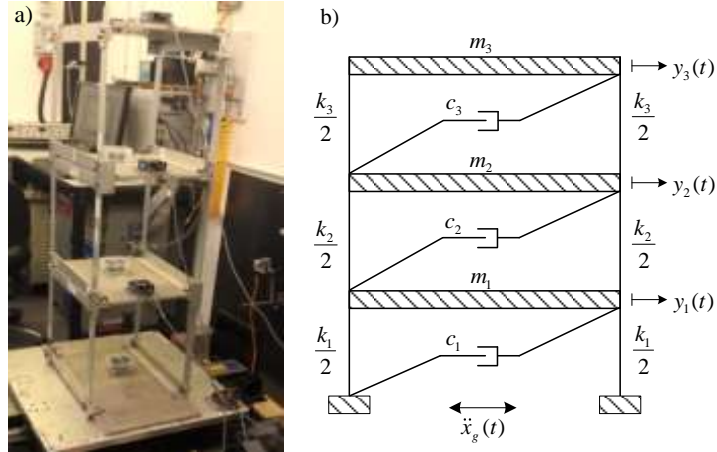

**Fig. 14.** (a) three-story frame tested on a shaking-table (b) 3-DOF linear structural model

Time-history acceleration responses of the 1st and 3rd stories were taken as the measured quantities using $nN_o = 12000$ number of data points for the inference. The MMTE kernel with additive GWN is used for describing the discrepancy between the measured and model responses. Similar to the above example, the proper order of the kernel function is found by comparing the BIC score for $m = \{1, 2, 3\}$. This example is a 3-DOF structure, so only up to three dynamical modes can be observed in the model response. Therefore, the order of the MMTE kernel cannot exceed three. As compared in Fig. 15, the BIC scores suggest that $m = 3$ is a good choice. Thus, the unknown parameters of the kernel function can be included into $\boldsymbol{\phi} = [\sigma_{f,1}^2\ \omega_1\ \ell_1^{-2}\ \sigma_{f,2}^2\ \omega_2\ \ell_2^{-2}\ \sigma_{f,3}^2\ \omega_3\ \ell_3^{-2}\ \sigma_n^2]^T$.

The MPV of the parameters are calculated through Algorithm 1, and the results are reported in Tables 3-4. The MPV of the structural parameters are compared with their reference values in Table 3, verifying the validity of the results. The posterior uncertainty of the stiffness and damping parameters is approximated using Laplace approximation, and the standard deviation is provided in Table 3. Additionally, in Table 4, the optimal values of the kernel parameters are provided. It can be observed that the variances $\sigma_{f,i}^2$'s are much larger than $\sigma_n^2$, showing that a large portion of the uncertainty is captured by the kernel covariance function and not the GWN additive component. The correlation lengths $\ell_i$'s appear to be between $1.0\ \text{s} < \ell_i < 1.25\ \text{s}$, and the frequencies of the MMTE kernel function ($\omega_i$'s) seem to be very close to those of the structure, reported in [63] as {27.57, 81.36, 118.20 rad/s}.

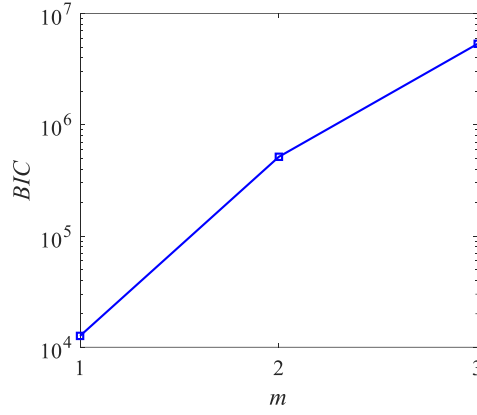

**Fig 15.** Selection of the order of the kernel covariance function

**Table 3.**
Posterior distribution of the structural parameters approximated using Algorithm 1

| Structural Parameters | Nominal values | MPV | SD |
|---|---|---|---|
| $k_1$ (kN/m) | 17.49 | 17.86 | 0.16 |
| $k_2$ (kN/m) | 25.66 | 24.20 | 0.53 |
| $k_3$ (kN/m) | 24.79 | 26.00 | 0.38 |
| $\zeta_1$ (%) | 2.39 | 1.89 | 0.73 |
| $\zeta_2$ (%) | 0.87 | 0.34 | 0.15 |
| $\zeta_3$ (%) | 0.65 | 1.30 | 0.44 |

**Table 4.**
Optimal values of the parameters of the MMT kernel combined with GWN

| Parameters | $\sigma_{f,1}^2$ | $\sigma_{f,2}^2$ | $\sigma_{f,3}^2$ | $1/\ell_1^2$ | $1/\ell_2^2$ | $1/\ell_3^2$ | $\omega_1$ | $\omega_2$ | $\omega_3$ | $\sigma_n^2$ |
|---|---|---|---|---|---|---|---|---|---|---|
| MPV | 0.062 | 0.027 | 0.042 | 0.647 | 0.727 | 0.996 | 114.47 | 79.66 | 26.11 | $5\times10^{-5}$ |

Fig. 16 compares the measured acceleration response of the first floor with the uncertainty bounds obtained by the GP framework. Fig. 17 shows the uncertainty bounds along with the structural model discrepancy. As can been observed, the uncertainty bounds gradually expands as we deviate from the training segment ended at $t=10$s, but it reaches a stationary regime after $t=15$s. However, care should be taken as the estimated mean of the model discrepancy is reliably estimated only for a short period of time, [10-10.5s], lasting only after half a length-scale ($0.5 \text{ s} < \ell_i/2 < 0.75 \text{ s}$). Unlike the predicted mean, the indicated uncertainty bound almost entirely covers the measured response and can reflect a reasonable degree-of-confidence.

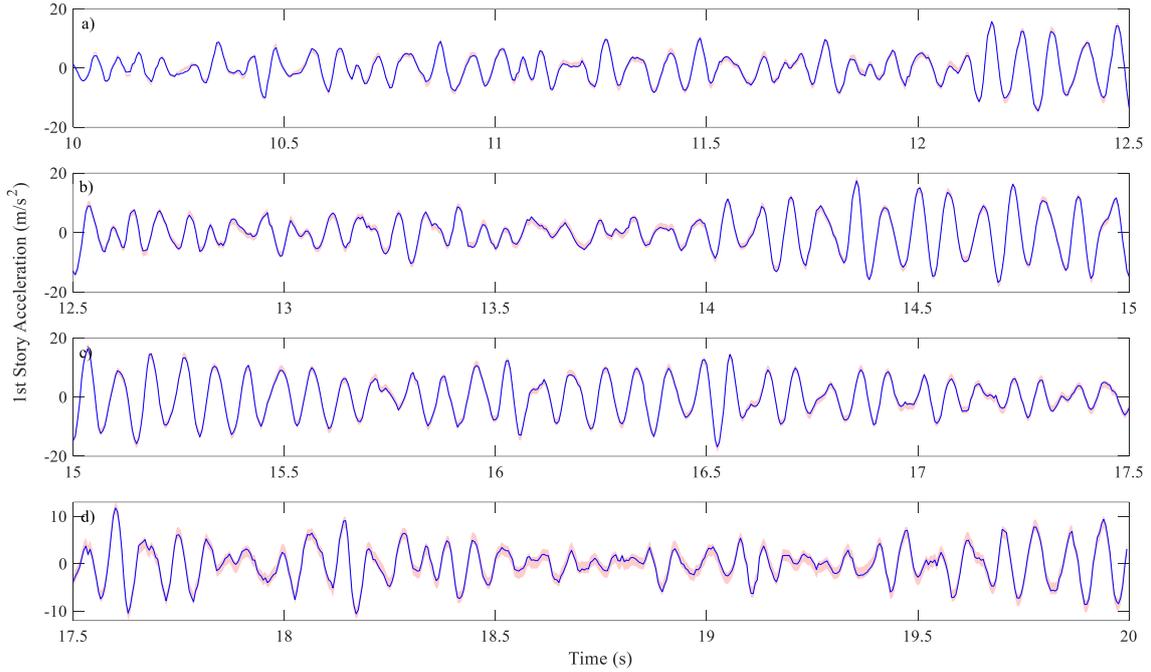

**Fig. 16.** Prediction of the at the 1<sup>st</sup> floor acceleration response (the shaded area shows ±3SD uncertainty bound; the blue line shows the measured response)

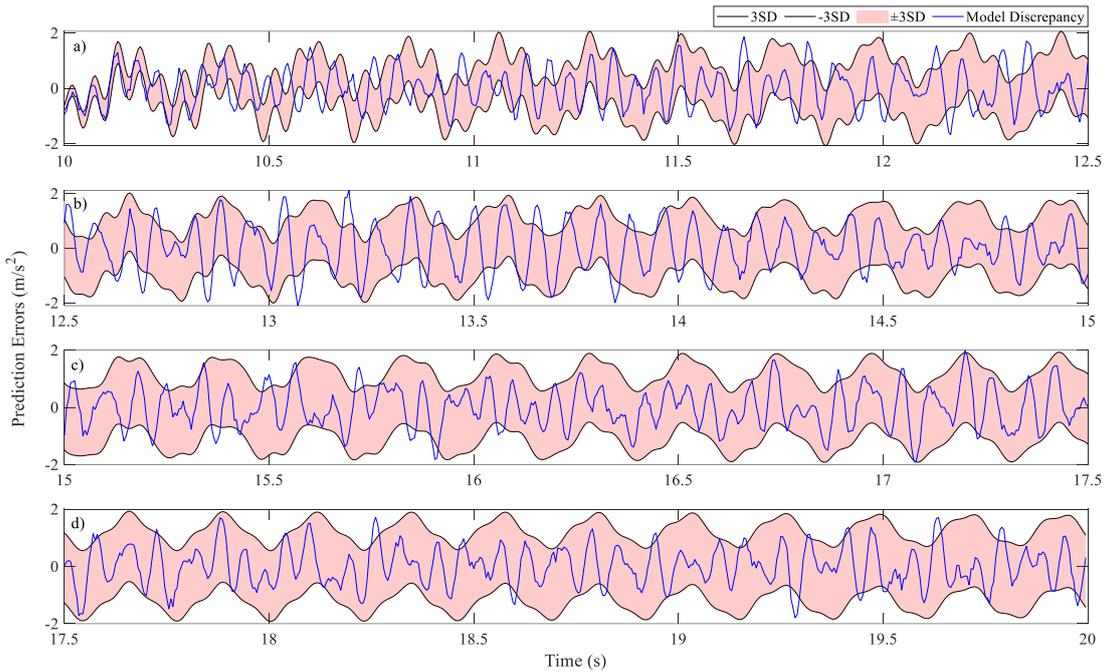

**Fig. 17.** Prediction of the at the 1<sup>st</sup> floor acceleration response

## 7. Conclusions and Future Works

A novel Bayesian framework is proposed for structural model updating, response predictions, and missing samples reconstruction using GP models. Kernel functions are introduced for modeling the correlation of prediction errors. Potential merits of these kernels are discussed in terms of how well the GP model compensates for model discrepancies. The choice of kernel functions is directed toward Bayesian model class selection, and new formulations are offered to incorporate the accuracy of response predictions, the fitting accuracy of training data, and model complexity. However, the computational cost of training the GP models turns out to be challenging mainly due to the rise in the number of unknown parameters, as well as the increase of the size of covariance matrix with the number of data points. Thus, a few computational strategies are developed for model inference and predictions using Laplace approximation and MCMC sampling techniques. The computational cost of calculating the inverse and the determinant of the covariance matrix is also addressed using a truncated spectral representation. Finally, the proposed framework is tested and verified via numerical and experimental examples. The modelling errors considered in the numerical examples were introduced through misspecification of damping ratios. However, the proposed method can capture errors created due to other sources of uncertainty, including boundary conditions, thermal effects, etc. Overall, the following conclusions can be drawn from the presented results:

- The proposed framework provides accurate parameter and response estimates, compensating for the structural model discrepancy. It also delivers reasonable response uncertainties, covering potential prediction errors.
- Based on the sample autocorrelation function and the spectral density of the model discrepancy (prediction errors), it is seen that the errors can be approximated as a few dominant mixing cosine functions.
- The MMTE kernel functions offer the best prediction performance compared to other choices. This finding is supported by comparing the relative plausibility of different kernels calculated based on the posterior probability, as well as the quality of response predictions in held-out data sets.
- The number of modes incorporated into the kernel function has a large effect on the parameter estimates and response predictions. Therefore, it is advisable to include the most dominant modes in the signal of the interest and set the frequencies of the MMTE kernel function initially at the modal frequencies of the structure.
- The BIC scores are observed to provide reliable results for selecting a proper model order for the MMTE covariance function.

In this paper, the proposed MMTE function is shown to be promising for dealing with linear structural models when stationary kernel functions are justifiable. Although the proposed method requires the knowledge of input forces, the concept seems to remain relevant to problems with unknown input. In this regard, one possible approach is to integrate the proposed method within the coupled input-state-parameter estimation methods using Bayesian filtering techniques, e.g., [64]. Thus, our future works will generalize this framework by considering nonlinearities and unknown input forces, as well as implementing non-stationary classes of covariance functions.

**Acknowledgements**
Financial support from the Hong Kong Research Grants Council under grants 16212918 and 16211019 is gratefully appreciated.